\documentclass[extra]{gji}

\usepackage{graphicx}
\usepackage{hyperref}
\usepackage{amsfonts}

\usepackage[T3,OT1]{fontenc}
\DeclareSymbolFont{tipa}{T3}{cmr}{m}{n}
\DeclareMathAccent{\invbreve}{\mathalpha}{tipa}{16}

\date{Accepted 2020 September 23. Received 2020 September 9; in original form 2020 July 13}

\begin{document}
\title[Iterative Marchenko scheme for imperfectly sampled data]{Adaptation of the iterative Marchenko scheme for imperfectly sampled data}
\author[J. van IJsseldijk and K. Wapenaar]{Johno van IJsseldijk$^1$\thanks{Email: \href{mailto:J.E.vanIJsseldijk@tudelft.nl}{J.E.vanIJsseldijk@tudelft.nl}} and Kees Wapenaar$^{1}$  \\
$^1$ Department of Geoscience and Engineering, Delft University of Technology, Delft, The Netherlands }

\maketitle

\begin{summary}
The Marchenko method retrieves the responses to virtual sources in the Earth's subsurface from reflection data at the surface, accounting for all orders of multiple reflections. The method is based on two integral representations for focusing- and Green's functions. In discretized form, these integrals are represented by finite summations over the acquisition geometry. Consequently, the method requires ideal geometries of regularly sampled and co-located sources and receivers. Recently new representations were derived, which handle imperfectly sampled data. These new representations use point-spread functions (PSFs) that reconstruct results as if they were acquired using a perfect geometry. Here, the iterative Marchenko scheme is adapted, using these new representations, to account for imperfect sampling. This new methodology is tested on a 2D numerical data example. The results show clear improvement of the proposed scheme over the standard iterative scheme. By removing the requirement for perfect geometries, the Marchenko method can be more widely applied to field data.
\end{summary}

\section{Introduction}

Seismic surveys are generally concerned with targets in the Earth's subsurface. However, structures in the overburden can distort the response of deeper targets. Ideally, all overburden structures and their multiple reflections should entirely be removed from the data, leaving only the response of the desired deeper targets. This can be achieved by redatuming the reflection response measured at the surface to a new datum plane below the overburden. The data-driven Marchenko method allows for the placement of virtual sources anywhere inside the subsurface, while accounting for all orders of multiples of the overburden \citep{broggini2012focusing,wapenaar2014marchenko,slob2014seismic}. Thereafter, the receivers can be moved to the same datum plane by a multidimensional deconvolution. Thus, Marchenko redatuming effectively shifts the response from the surface to a new datum inside the medium, and fully removes all interactions with the shallower structures. \\
Although the method has been successfully applied to real data  \citep[e.g.][]{ravasietal2016,staringetal2018}, several constraints still limit the usefulness of the method. 
Marchenko redatuming is based on two integral representations. These coupled equations can be solved by direct inversion \citep{vanderneut2015inversion} or by iterative substitution \citep{thorbecke2017implementation}. In practice, the infinite integrals are replaced by summations over the finite acquisition geometry. This requires regularly sampled and collocated sources and receivers in order to retrieve proper, uncontaminated responses. On the contrary, non-perfect geometries can have a significant effect on the Marchenko results \citep{peng2019effects,staring2019interbed}. Most authors, therefore, assume ideal acquisition geometries when using the Marchenko method, thus avoiding the limitations arising from imperfect sampling. However, this restriction should ideally be relaxed or even removed, allowing for broader application of the method on field data. \\
\citet{peng2019subsamp} consider the effects of different sub-sampling and integration scenarios. Two main effects are identified. First, when the sub-sampling and integration occur over the same dimension, the focusing- and Green's functions get distorted but remain well-sampled. Second, in the situation of sub-sampling and integration over different dimensions, the focusing- and Green's functions are accurate for the non-zero traces but contain spatial gaps. In the case of irregular sampling, the second effect can partly be removed by using a sparse inversion of the Marchenko equations, outputting well-sampled focus functions and sub-sampled Green's functions \citep{ravasi2017rayleigh,haindl2018sparsity}. On the other hand, \citet{wapenaar2019} introduce new representations for focusing- and Green's functions, that are distorted by imperfect sampling and integration over the same dimension. Inverting these representations involves a multidimensional deconvolution with novel point-spread functions (PSFs) to deblur the distorted focusing- and Green's functions.
These representations are then verified on analytically modeled focusing functions, that have been derived from decomposed wave-field propagators and scattering coefficients. However, in real scenarios these functions are unavailable and have to be derived from the coupled Marchenko equations. \\
In this paper we explore how we can integrate the new representations for irregularly sampled data into the iterative Marchenko scheme. First, the theory of deblurring the Marchenko equations with PSFs is reviewed. Next, the paper discusses the required changes to apply PSFs in the iterative scheme. Then, we present an altered version of the iterative scheme, that allows for imperfectly sampled data. The performance of the newly developed scheme is then tested on numerical examples. Although the results are promising, the stability of the scheme is uncertain and only assured for certain subsurface models. The last part of the paper, therefore, presents a modified scheme with greater stability, which is less susceptible to subsurface conditions.

\section{Green's function representations}
\newcommand{\xs}[1]{\textbf{x}_{#1}}
This section reviews briefly the theory of the Green's function representations that are the basis for the  Marchenko method. For a more elaborate derivation the reader is referred to \citet{wapenaar2014marchenko} and \citet{slob2014seismic}. As starting point, imagine an inhomogeneous lossless subsurface bounded by transparent acquisition surface $\mathbb{S}_0$. The reflection response at this surface is given by $R(\xs{R},\xs{S},t)$, with $\xs{S}$ a dipole source, $\xs{R}$ monopole receivers, and $t$ denoting the time. In this paper we investigate how to account for irregularly sampled sources. Via reciprocity,  $\xs{R}$ can be interpreted as a monopole source and $\xs{S}$ as a dipole receiver.  Hence, the method developed in this paper can also be used to account for irregularly sampled receivers. We define the focal depth at surface $\mathbb{S}_A$, on which the virtual receivers are located. These virtual receivers are used to observe the up- and down-going Green's functions: $G^-(\xs{A},\xs{R},t)$ and $G^+(\xs{A},\xs{R},t)$, respectively. Here, $\xs{A}$ is the location of the virtual receivers at the focal depth. For the definition of the focusing functions, the medium is truncated below the focal depth, resulting in a medium that is inhomogeneous between $\mathbb{S}_0$ and $\mathbb{S}_A$, and homogeneous above and below these surfaces. In this medium we define a downgoing focusing function $f^+_1(\xs{S},\xs{A},t)$, which, when injected from the surface, focuses at the focal depth $\mathbb{S}_A$ at $\xs{A}$. Moreover, $f^-_1(\xs{R},\xs{A},t)$ is the upgoing response of the truncated medium as measured at the surface, known as the upgoing focusing function. 
These ideas can be combined in two integral equations, as follows \citep{wapenaar2014marchenko,slob2014seismic}:
\[
G^-(\xs{A},\xs{R},t) + f^-_1(\xs{R},\xs{A},t) = 
\]
\begin{equation}
\label{eqn:mar1}
\hspace*{1cm} \int_{\mathbb{S}_0} R(\xs{R},\xs{S},t) * f^+_1(\xs{S},\xs{A},t)d\xs{S},
\end{equation}

\[
G^+(\xs{A},\xs{R},t) - f^+_1(\xs{R},\xs{A},-t) = 
\]
\begin{equation}
 \label{eqn:mar2}
\hspace*{1cm} -\int_{\mathbb{S}_0} R(\xs{R},\xs{S},t) * f^-_1(\xs{S},\xs{A},-t)d\xs{S}.
\end{equation}

\noindent
The asterisk in these equations denotes a temporal convolution. For acoustic media, the focusing- and Green's functions on the left-hand side are separable in time by a windowing function. In practice, the infinite integrals on the right-hand side are approximated by finite sums over the available sources. For the RHS of \autoref{eqn:mar1} this yields: 
\begin{equation}
\label{eqn:dis1}
\sum_{i} R(\xs{R},\xs{S}^{(i)},t) * f^+_1(\xs{S}^{(i)},\xs{A},t)*S(t) , 
\end{equation}
and for the RHS of \autoref{eqn:mar2}:
\begin{equation}
\label{eqn:dis2}
-\sum_{i} R(\xs{R},\xs{S}^{(i)},t) * f^-_1(\xs{S}^{(i)},\xs{A},-t)*S(t) , 
\end{equation}
\noindent
where $i$ denotes the source position and $S(t)$ the source signature. When the reflection response is not well sampled, these summations cause distortions in the responses on the left-hand sides of \autoref{eqn:mar1} and \ref{eqn:mar2}. \\

\section{Point-spread functions}
\citet{wapenaar2019} introduce point-spread functions (PSFs) to correct for imperfect sampling. These PSFs exploit the fact that the ideal downgoing focusing function is the inverse of the transmission response. A convolution of the focusing function with the transmission response $T$ should, therefore, give a bandlimited delta pulse in space and time:
\[
\delta(\xs{H,A}'-\xs{H,A})\delta(t)= 
\]
\begin{equation}
\label{eqn:TID}
\hspace*{1cm} \int_{\mathbb{S}_0} T(\xs{A}',\xs{S},t)* f^+_1(\xs{S},\xs{A},t)d\xs{S}.
\end{equation}
An alternative form with integration over the focal depth is given by: 
\[
\delta(\xs{H,S}-\xs{H,S}')\delta(t)= 
\]
\begin{equation}
\label{eqn:TID2}
\hspace*{1cm} \int_{\mathbb{S}_A}  f^+_1(\xs{S},\xs{A},t) * T(\xs{A},\xs{S}',t)d\xs{A}.
\end{equation}
However, for imperfectly sampled data this delta pulse gets blurred. This blurring quantifies the imperfect sampling, as follows:
\[
\Gamma^+_1(\xs{A}',\xs{A},t)= 
\]
\begin{equation}
\label{eqn:gammaplus}
\hspace*{1cm} \sum_{i} T(\xs{A}',\xs{S}^{(i)},t)* f^+_1(\xs{S}^{(i)},\xs{A},t)*S(t).
\end{equation}

\noindent
Here $\Gamma^+_1$ is the downgoing PSF. Similarly, a quantity $Y_1$ is defined as the inverse of the time-reversed, upgoing focusing function:
\[
\delta(\xs{H,A}'-\xs{H,A})\delta(t)= 
\]
\begin{equation}
\label{eqn:YID}
\hspace*{1cm} \int_{\mathbb{S}_0} Y_1(\xs{A}',\xs{S},t) *f^-_1(\xs{S},\xs{A},-t)d\xs{S},
\end{equation}
or alternatively:
\[
\delta(\xs{H,S}-\xs{H,S}')\delta(t)= 
\]
\begin{equation}
\label{eqn:YID2}
\hspace*{1cm} \int_{\mathbb{S}_A}  f^-_1(\xs{S},\xs{A},-t) * Y_1(\xs{A},\xs{S}',t) d\xs{A},
\end{equation}
Again, the irregular sampling will result in a blurring of the delta pulse on the LHS of \autoref{eqn:YID}. The convolution to quantify the upgoing PSF ($\Gamma^-_1$) then becomes:
\[
\Gamma^-_1(\xs{A}',\xs{A},t)= 
\]
\begin{equation}
\label{eqn:gammamin}
\hspace*{1cm} \sum_{i} Y_1(\xs{A}',\xs{S}^{(i)},t) *f^-_1(\xs{S}^{(i)},\xs{A},-t)*S(t).
\end{equation}

\begin{figure}
\centering
\includegraphics[width=0.48\textwidth]{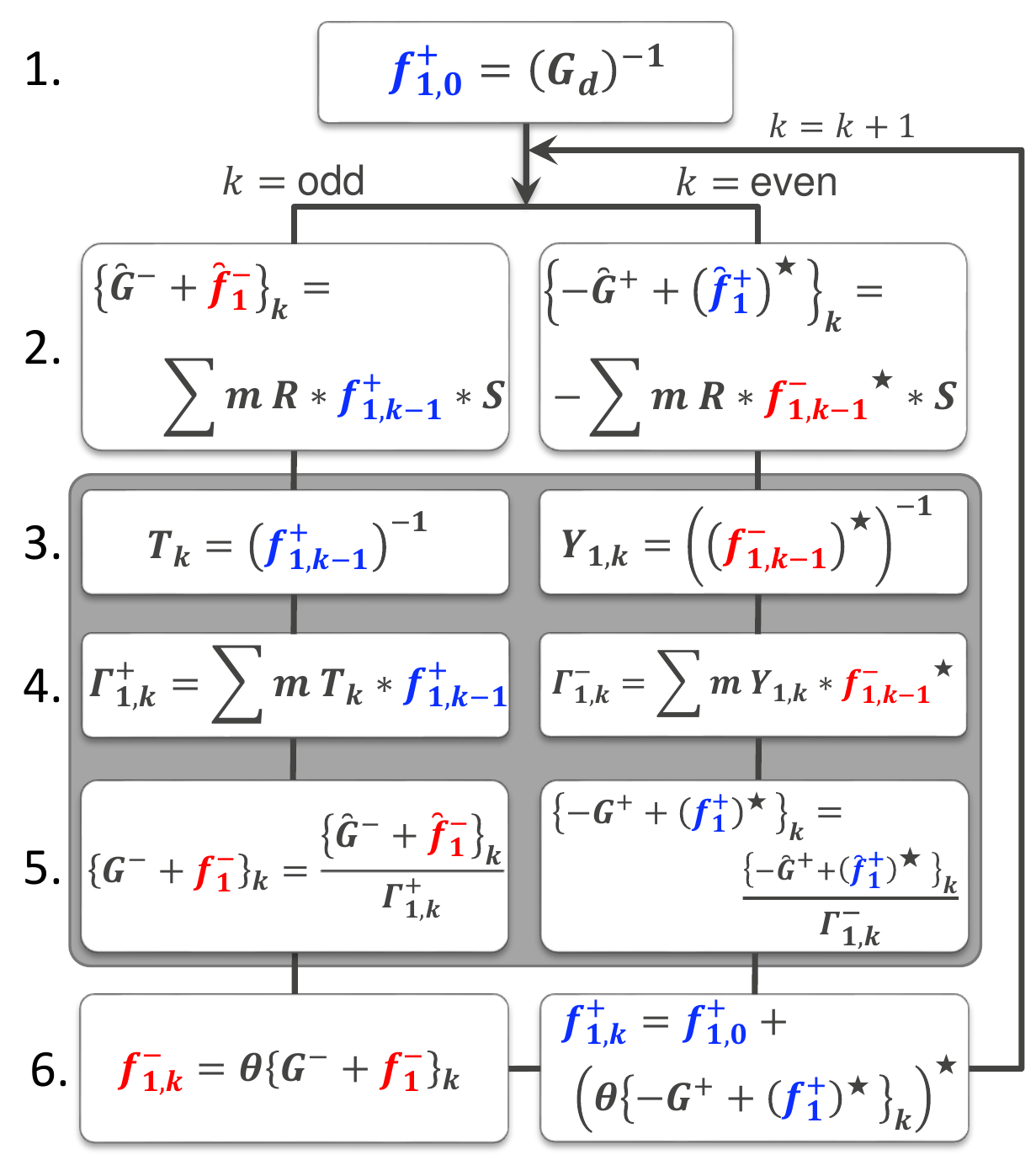}
\caption{Flowchart with the proposed iterative Marchenko scheme, steps 3 to 5 account for imperfectly sampled data. Here $f$ and $G$ represent the focusing- and Green's functions, respectively, $S$ is the source signature. $m$ is a masking operator, that kills all the traces with missing sources. $k$ denotes the iteration number. The arch over a symbol denotes that the response is contaminated by the imperfect sampling, the superscript star denotes time-reversal. The inline asterisks denote convolutions or correlations, which are then summed over the imperfectly sampled sources. Finally, $\mathbf{\theta}$ is the time-windowing operator.}
\label{is_flowchart}
\end{figure}

\noindent
Once again, in the case of perfect sampling this PSF would be equal to a bandlimited delta pulse in space and time. Note that this inverse ($Y_1$) is not necessarily stable, because $f^{-}_1$ is a reflection response. On the contrary, $f^{+}_1$ is more stable and more likely to be invertible (i.e. in the limiting case of a 1D medium it is a minimum-phase function, which is always invertible). This will be elaborated upon in the discussion section. \\
Next, \citet{wapenaar2019} apply these newly acquired PSFs to \autoref{eqn:mar1} and \ref{eqn:mar2}, respectively. For both sides of \autoref{eqn:mar1} we employ the operator $\int_{\mathbb{S}_A} \{\cdot\} * \Gamma^+_1(\xs{A}',\xs{A},t) d\xs{A}'$, whereas both sides of \autoref{eqn:mar2} require the use of the operator $\int_{\mathbb{S}_A} \{\cdot\} * \Gamma^-_1(\xs{A}',\xs{A},t) d\xs{A}'$. The resulting equations can be further simplified using equations \ref{eqn:TID2}, \ref{eqn:gammaplus}, \ref{eqn:YID2} and \ref{eqn:gammamin} to derive two new representations for irregularly sampled data:
\[
\invbreve{G}^-(\xs{A},\xs{R},t) + \invbreve{f}^-_1(\xs{R},\xs{A},t) = 
\]
\begin{equation}
\label{eqn:mar3}
\hspace*{1cm} \sum_{i} R(\xs{R},\xs{S}^{(i)},t) * f^+_1(\xs{S}^{(i)},\xs{A},t)*S(t),
\end{equation}
\[
\invbreve{G}^+(\xs{A},\xs{R},t) - \invbreve{f}^+_1(\xs{R},\xs{A},-t) = 
\]
\begin{equation}
\label{eqn:mar4}
\hspace*{1cm}-\sum_{i} R(\xs{R},\xs{S}^{(i)},t) * f^-_1(\xs{S}^{(i)},\xs{A},-t)*S(t),
\end{equation}

\noindent 
with:
\[
\invbreve{G}^{\pm}(\xs{A},\xs{R},t) = 
\]
\begin{equation}
\hspace*{1cm}\int_{\mathbb{S}_A} G^{\pm}(\xs{A}',\xs{R},t) * \Gamma_1^{\mp}(\xs{A}',\xs{A},t) d\xs{A}',
\end{equation}
and
\[
\invbreve{f}_1^{\pm}(\xs{R},\xs{A},\mp t) = 
\]
\begin{equation}
\hspace*{1cm}\int_{\mathbb{S}_A} f^{\pm}_1(\xs{R},\xs{A}',\mp t) * \Gamma_1^{\mp}(\xs{A}',\xs{A},t) d\xs{A}'.
\end{equation}
Equations \ref{eqn:mar3} and \ref{eqn:mar4} have two interesting features. First, the right-hand sides are now the same as Equations \ref{eqn:dis1} and \ref{eqn:dis2}. Second, the responses on the left-hand sides now contain the PSFs, which apply a blurring effect to each response. Note that the imperfectly sampled data can now be deblurred by a multidimensional deconvolution (MDD) with the PSFs, assuming these PSFs are known.

\begin{figure}
\centering
\includegraphics[width=.48\textwidth]{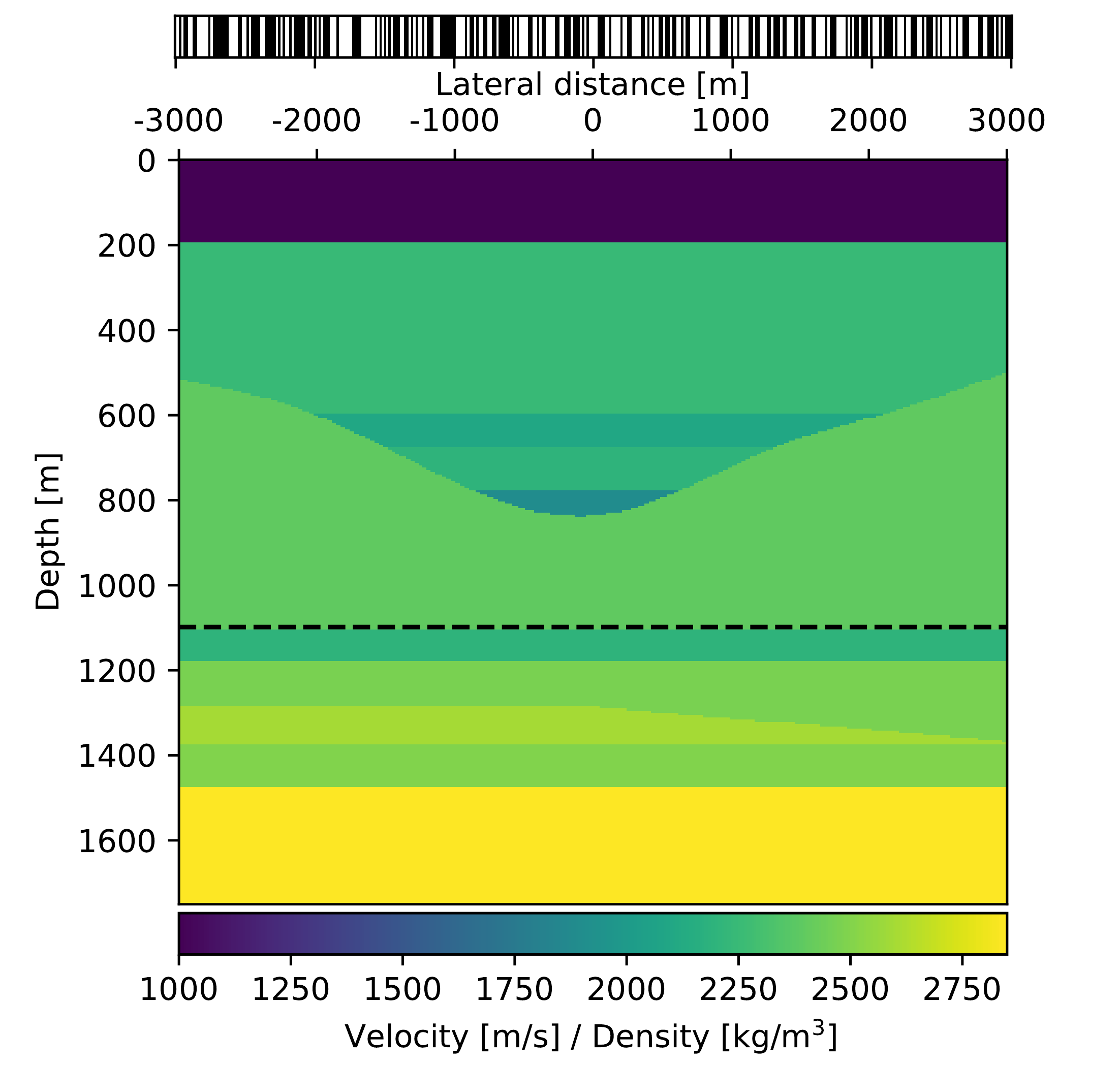}
\caption{Model used in the numerical irregular sampling experiment, the dashed line shows the focal level. The barcode shows the irregular sampling, with the white spaces denoting the excluded sources.}
\label{is_model}
\end{figure}

\section{Iterative Marchenko scheme}

\citet{wapenaar2019} verify the representations in equations \ref{eqn:mar3} and \ref{eqn:mar4}, using analytically modelled focusing functions (i.e. both the reflection response and focusing functions on the RHS of the equations are known). In practice, these focusing functions are unknown, and have to be retrieved from the Marchenko equations. This can be achieved iteratively or by inversion of the Marchenko equations. Here, we aim to integrate the representations for imperfectly sampled data with the iterative approach \citep{thorbecke2017implementation}.   \\
\autoref{is_flowchart} shows the proposed iterative Marchenko scheme, which corrects for imperfect sampling in each iteration $k$. The first step is to estimate the initial downgoing focusing function ($f^+_{1,0}$). Traditionally, this is approximated by the time-reversal of the direct arrival of the Green's function. However, to ensure that the convolution of the transmission response and downgoing focusing function gives a delta pulse in space and time with the correct amplitudes, the proposed scheme inverts the direct arrival in step 1:
\begin{equation}
\label{eqn:inif1p}
f^+_{1,0} (\xs{S},\xs{A},t) \approx G^{inv}_d(\xs{S},\xs{A},t).
\end{equation}
In practice this inversion is achieved by a least-squares-based inversion in the frequency domain, 
 where for each frequency slice a band-limited identity matrix (delta pulse) is divided by $G_d$ to find $f^+_{1,0}$. Note that this approach requires a matrix with a size equal to the number of shots multiplied with the number of focal points, and can, therefore, not be done for a single focusing point. \\
The next step computes the focusing- and Green's function by a convolution or correlation  for the odd or even iterations, respectively. The odd iterations are computed according to \autoref{eqn:mar3}, where the downgoing focusing function on the RHS is retrieved from the initial condition for the first iteration or from the previous iteration for subsequent iterations. Similarly, the even iterations use the upgoing focusing functions from the previous iteration in the correlation with the reflection response, as shown in \autoref{eqn:mar4}. Note, for well-sampled data the computed focusing- and Green's functions in this step are free of distortions, therefore the resulting focusing- and Green's functions are equal to these functions in the standard scheme:
\[
\lbrace\invbreve{G}^\pm (\xs{A},\xs{R},t) \mp \invbreve{f}^{\pm}_1(\xs{R},\xs{A},\mp t)\rbrace _k = 
\]
\begin{equation}
\hspace*{1cm} \lbrace G^\pm (\xs{A},\xs{R},t) \mp f^{\pm}_1(\xs{R},\xs{A},\mp t)\rbrace _k.
\end{equation}
In this case steps 3 to 5 are redundant and can be omitted, this indeed reduces the proposed scheme to the standard iterative Marchenko scheme. \\
For irregularly sampled reflection data, steps 3 to 5 are introduced. The first objective is to find an estimate of the transmission response and quantity $Y_1$ for odd and even iterations, respectively. Since these responses are defined as the inverse of the focusing functions, they can be obtained by inversion of the following equations:
\[
\delta(\xs{H,A}'-\xs{H,A})\delta(t)= 
\]
\begin{equation}
\label{eqn:Tk}
\hspace*{1cm} \int_{\mathbb{S}_0} T_k(\xs{A}',\xs{S},t)* f^+_{1,k-1}(\xs{S},\xs{A},t)d\xs{S},
\end{equation}
and
\[
\delta(\xs{H,A}'-\xs{H,A})\delta(t)=
\]
\begin{equation}
\label{eqn:Yk}
\hspace*{1cm} \int_{\mathbb{S}_0} Y_{1,k}(\xs{A}',\xs{S},t) *f^-_{1,k-1}(\xs{S},\xs{A},-t)d\xs{S}.
\end{equation}
$T_k$ in \autoref{eqn:Tk} denotes the estimated transmission response for each odd iteration $k$, and $f^+_{1,k-1}$ is the downgoing focusing function computed in the former iteration $k-1$. \autoref{eqn:Yk} computes an approximation of the quantity $Y_{1,k}$ for each even iteration, based on the upgoing focusing function from the preceding iteration. Note that both the up- and downgoing focusing functions are deblurred, and free of distortions from the imperfect sampling. The two integral representations are, therefore, evaluated over a regular grid (i.e. as if no sources are missing). Next, the PSFs have to be computed, using the estimates of $T$ and $Y_1$ (step 4 in \autoref{is_flowchart}). Analogous to \autoref{eqn:gammaplus}, the downgoing PSF for each odd iteration is retrieved by evaluating the convolution of $T_k$ and $f^+_{1,k-1}$ over the irregular sampled sources. For the even iterations we consider the correlation of $Y_{1,k}$ and $f^-_{1,k-1}$, as in \autoref{eqn:gammamin}. Subsequently, in step 5 the distorted focusing- and Green's functions, from step 2 of the scheme, are deblurred by a multidimensional deconvoltion with the PSFs. Similar as in \autoref{eqn:inif1p}, this is effectively accomplished by a least-squares-based inversion, and once again it is done for all focal points simultaneously. Consequently, our scheme can not operate on individual focusing points, but rather multiple points are considered simultaneously. This is in contrast to the standard Marchenko scheme, which can operate on a single focusing point.
After the MDD, the resulting focusing- and Green's functions are reconstructed as if they were retrieved with well-sampled data. Finally, the last step separates the focusing function from the Green's function using a time-windowing operator ($\mathbf{\theta}$ in \autoref{is_flowchart}). This final step is identical to that in the standard Marchenko scheme. \\
Each iteration is initialized with a ``clean'' (i.e. deblurred) focusing function from the preceding iteration. This is required at the start of each iteration, otherwise the errors from the irregular sampled reflection data would accumulate. Therefore, steps 3 to 5 are enforced with every iteration.

\begin{figure*}
\centering
\includegraphics[width=\textwidth]{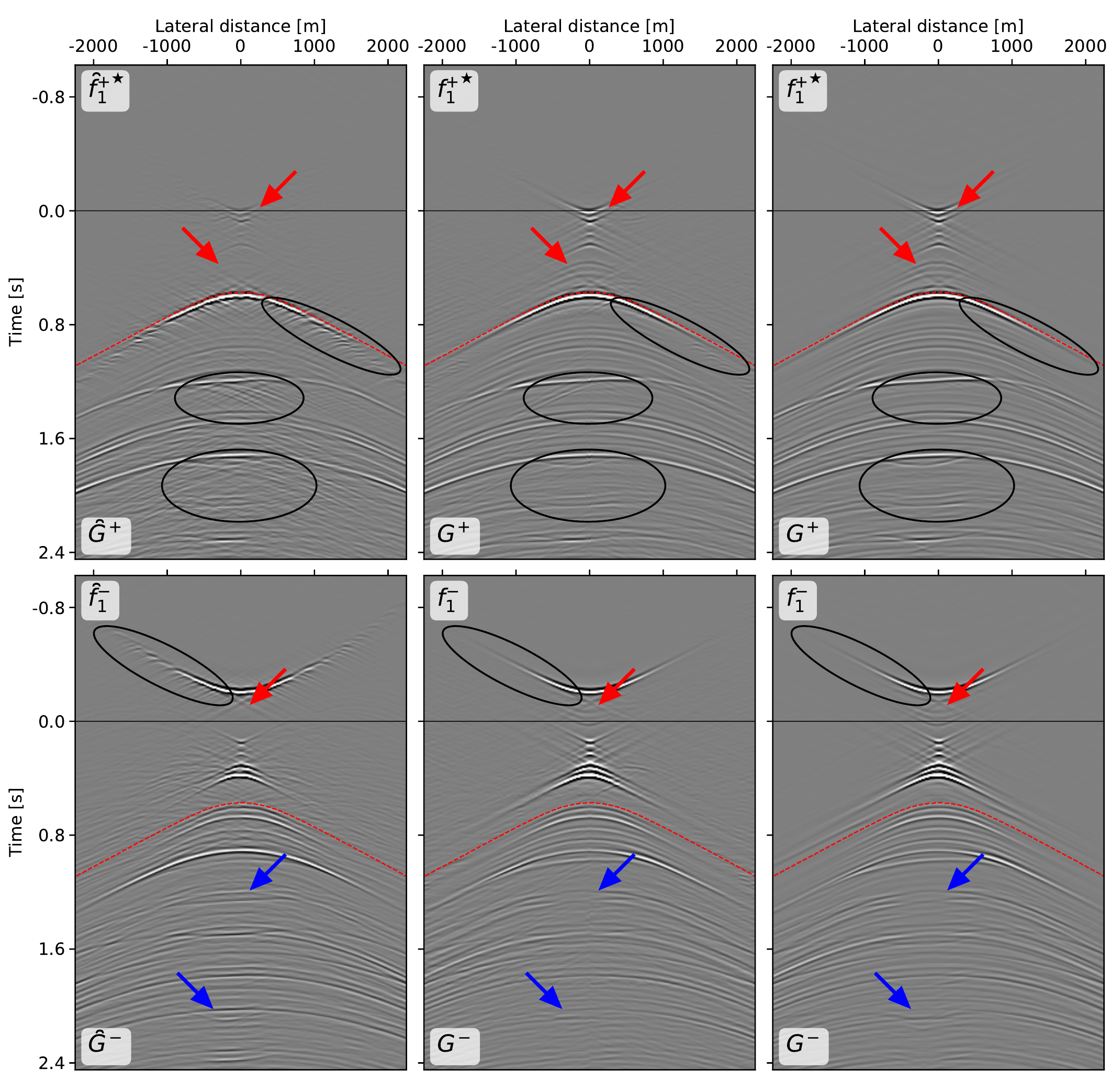} \\
\caption{The top row shows the time-reversed downgoing focusing function ($\{f_1^+\}^\star$) and downgoing Green's function ($G^+$), and the bottom row shows the upgoing focusing function ($f_1^-$) and upgoing Green's function ($G^-$), the star superscript denotes time-reversal. The dashed, red lines indicate the separation  between the focusing- and Green's functions. The left column shows the result of irregularly sampled data after 12 iterations of the standard Marchenko scheme. The middle column shows the results when using our scheme on the same data (\autoref{is_flowchart}), again 12 iterations are used. Finally, the 3rd column shows the reference result, obtained after 12 iterations of the standard Marchenko scheme with well-sampled data. Each panel is scaled with its maximum value. The arrows and ellipses show artifacts arising from the irregular sampling. Distortions caused by the irregular sampling are indicated with the ellipses. The red arrows show events that deviate in amplitude or are missing altogether. Finally, the blue arrows mark erroneous reflectors. }
\label{is_results}
\end{figure*}
\section{Numerical example}

The performance of the proposed scheme is tested on synthetic data. The 2D model for this test is shown in \autoref{is_model}. For convenience, the density and velocity parameters are chosen to be the same in each layer, but this is not required for successful application of the scheme. The observant reader will note the strong contrast in acoustic impedance between the top two layers of the model, at a depth of 200 meters. This contrast ensures that the inversion of $f^-_1$ for retrieving $Y_1$ is stable, because most of the energy gets concentrated at the early onsets of the reflection response. Note that this constraint on the subsurface model is significantly relaxed with a second scheme discussed later on in this paper. \\
The reflection response of the medium is modeled using a wavelet with a flat spectrum between 5 and 80 Hz. 
In total 601 sources and receivers are used with an initial spacing of 10 meters. For the irregular sampling 50\% of the sources are removed at random, as can be seen in the barcode plot in \autoref{is_model}. In practice, these sources are killed (i.e. set to 0), as opposed to being entirely removed from the reflection response. Next, the direct arrival of the Green's function between the focal depth and the Earth's surface is estimated in a smooth velocity model. As previously stated, the inverse of this direct arrival is used for the initial estimate of the upgoing focusing function, as opposed to the time-reversed version that is traditionally used. The reflection response and this initial estimate together are all the required inputs for the standard Marchenko scheme. Finally, for the fourth step of our proposed scheme the location of the sources (e.g. the barcode in \autoref{is_model}) is required. 

\begin{figure}
\centering
\includegraphics[width=.48\textwidth]{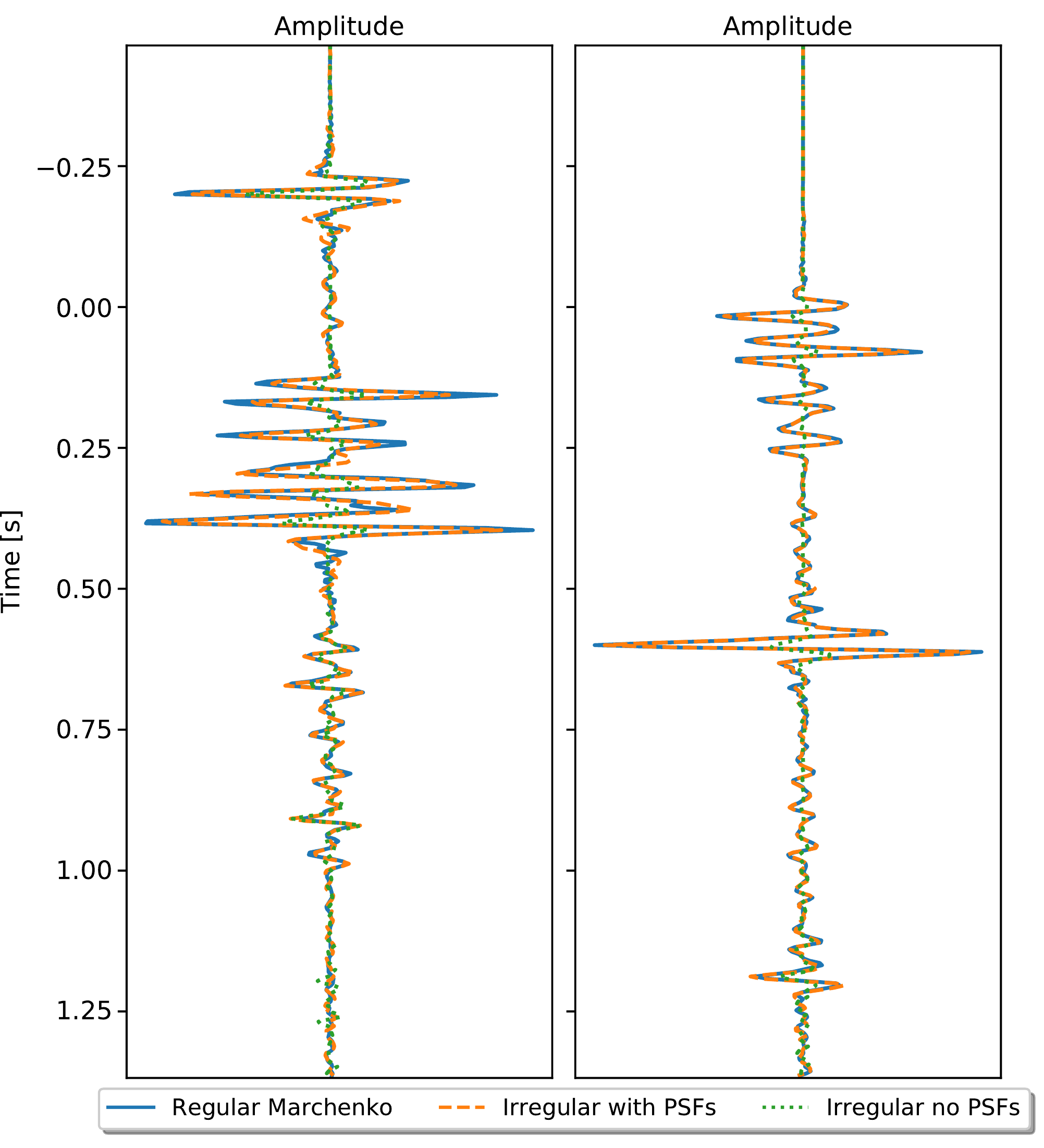}
\caption{Comparison of the amplitudes in the middle trace (at offset 0 m) of each panel in \autoref{is_results}. On the left are the time-reversed downgoing focusing function ($\{f_1^+\}^\star$) and downgoing Green's function ($G^+$). The upgoing focusing function ($f_1^-$) and upgoing Green's function ($G^-$) are shown on the right.}
\label{is_amplitudes}

\vspace{.4cm}
\includegraphics[width=0.48\textwidth]{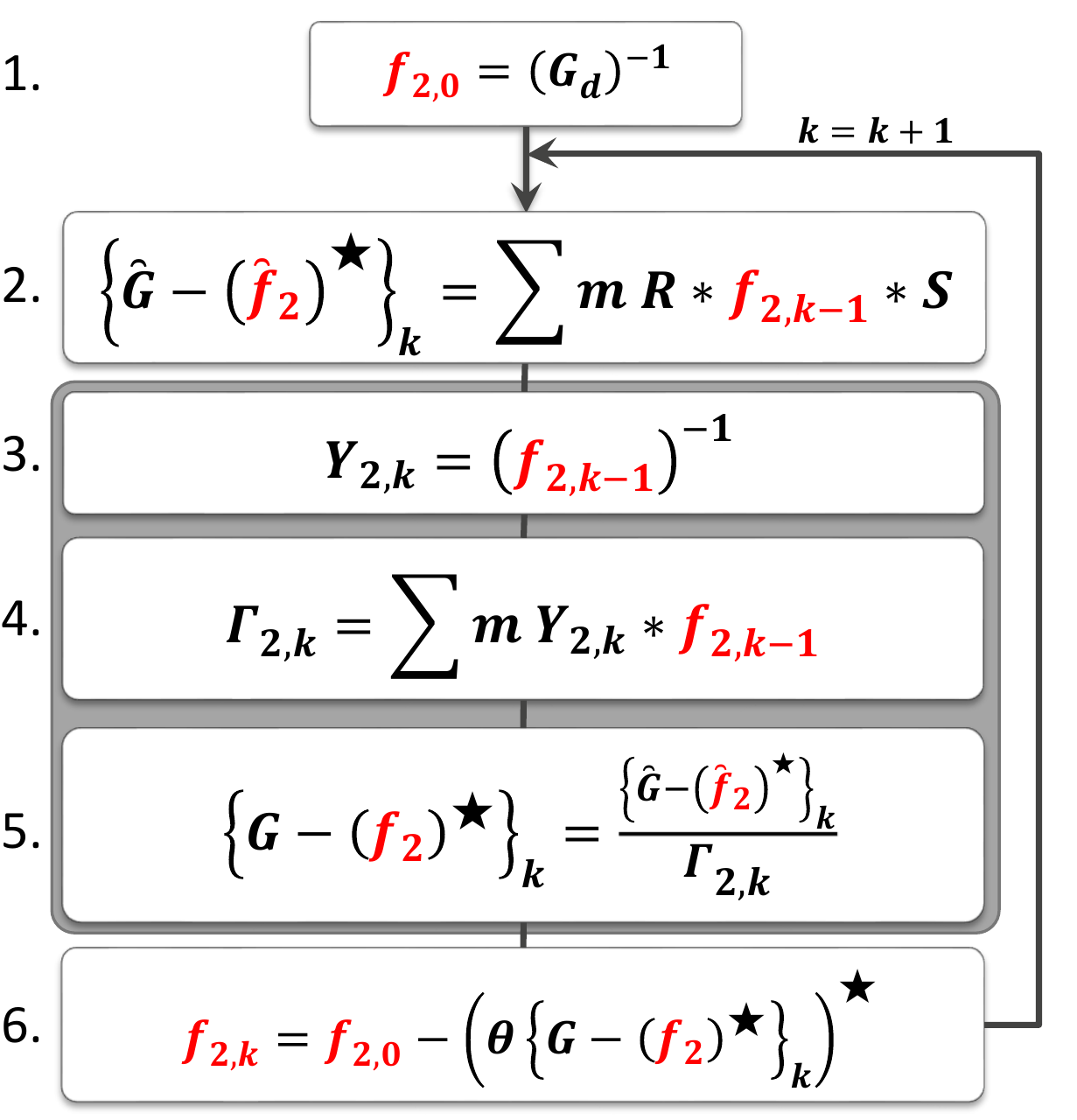}
\caption{Flowchart displaying the full wavefield Marchenko scheme, where steps 3 to 5 account for imperfectly sampled data. $f$ and $G$ represent the focusing- and Green's functions, respectively, and the arch denotes contamination by the imperfect sampling. $S$ is the source signature. $m$ is a masking operator, that kills all the traces with missing sources. $k$ is the iteration number, the superscript star denotes time-reversal, and the inline asterisks denotes a convolution. $\mathbf{\theta}$ is the time-windowing operator.}
\label{isf2_flowchart}
\end{figure}

\noindent
\autoref{is_results} shows the results of the numerical experiment, each column in the figure represents the results after 12 iterations using one of the three schemes. The first column shows the results where the standard Marchenko scheme is used with the irregularly sampled reflection data. Next, the middle column shows the results of the proposed scheme, again with irregularly sampled data. Finally, the last column displays a reference result, that was obtained by using the standard Marchenko scheme on reflection data without removing any sources. The red dashed line in the figure denotes the separation in time of the Green's functions below, and focusing functions above. In the case of irregular sampling in the standard scheme (as presented in the first column), three main artifacts can be identified. Firstly, clear distortions of some reflectors are observed, especially around the strong events.  These distortions are most noticeable of all artifacts, and obstruct later events in the downgoing Green's function ($\invbreve{G}^+$). The ellipses indicate some of these artifacts. Secondly, the amplitudes of some events are incorrect or the events are not reconstructed at all (as shown by the red arrows). For example, the downgoing focusing function ($\invbreve{f}^+$) is largely suppressed, as well as some events in the upgoing focusing function ($\invbreve{f}^-$). Lastly, some new and undesired reflectors are appearing in the results, especially at later times ($> 1.2$s) many of the reflectors in the upgoing Green's function ($\invbreve{G}^-$) are deviating from the reference result in the third column. Examples of such undesired reflectors are marked with the blue arrows. All these three types of artifacts are mostly removed by using the proposed scheme (middle column), and the results of this scheme show much more resemblance with the reference results. This implies that the proposed scheme both deblurs the results of irregular sampling effects, and also retrieves the amplitudes of the events more accurately. However, the method does introduce some of its own artifacts; as it introduces edge effects, especially at later times. These artifacts are introduced by the MDD of poorly sampled data with the PSFs, and they are suppressed by using directional FK-filters. \\
The amplitude reconstruction by the proposed scheme is further illustrated in \autoref{is_amplitudes}, where the middle trace of each panel from \autoref{is_results} is plotted. In \autoref{is_amplitudes} the results of the proposed scheme in orange quite closely match the reference results in blue, whereas the standard scheme fails to recover the correct amplitudes in the case of irregularly sampled reflection data (green line). This difference in amplitudes cannot simply be negated by scaling with a constant factor, because the error has a different magnitude at different times.

\section{A more stable alternative}

While the previous results show clear potential, the need for a stable inversion of $f^-_1$ imposes a large constraint on the subsurface models that are suitable for the method. This section, therefore, explores how this unstable inversion can be avoided. In order to achieve this equations \ref{eqn:mar1} and \ref{eqn:mar2} are combined into a single equation, that retrieves the full wavefield Green's functions between a focal point and surface, defined as follows:
\begin{equation}
\label{eqn:greensdecomp}
G(\xs{R},\xs{A},t) = G^+(\xs{A},\xs{R},t) + G^-(\xs{A},\xs{R},t).
\end{equation}
This gives a new representation for irregular sampling with the full wavefield Green's function:
\[
\invbreve{G}(\xs{R},\xs{A},t) - \invbreve{f}_2(\xs{A},\xs{R},-t) =
\]
\begin{equation}
\label{eqn:marnewf2}
\hspace*{1cm} \sum_{i} R(\xs{R},\xs{S}^{(i)},t) * f_2(\xs{A},\xs{S}^{(i)},t)*S(t),
\end{equation}
with: 
\begin{equation}
\label{eqn:deff2}
f_2(\xs{A},\xs{R},t) =  f^+_1(\xs{R},\xs{A},t) -  f^-_1(\xs{R},\xs{A},-t).
\end{equation}
The arches over the Green's and focusing functions in \autoref{eqn:marnewf2} denote convolution with a new PSF $\Gamma_2$:
\[
\Gamma_2(\xs{A}',\xs{A},t)= \\
\]
\begin{equation}
\label{eqn:gamma2plus}
\hspace*{1cm} \sum_{i} Y_2(\xs{A}',\xs{S}^{(i)},t)* f_2(\xs{A},\xs{S}^{(i)},t)*S(t).
\end{equation}
Here, $Y_2$ is the inverse of focusing function $f_2$. Note that this inverts a superposition of the downgoing and upgoing focusing functions, thereby avoiding the independent inversion of $f^-_1$ (see \autoref{eqn:Yk} ). A detailed derivation of equations \ref{eqn:marnewf2} to \ref{eqn:gamma2plus} is given in Appendix A.  \\
Next, we integrate these full wavefield equations into an iterative scheme, analogous to the integration of the decomposed equations shown before. An overview of this new iterative scheme is shown in \autoref{isf2_flowchart}. The scheme is initialized with an estimate for the first focusing function, which is again equal to the inverse of the direct arrival of the Green's function:
\begin{equation}
f_{2,0} (\xs{A},\xs{S},t) \approx G^{inv}_d(\xs{S},\xs{A},t).
\end{equation}
This initial focusing function is convolved with the subsampled reflection response (step 2 in \autoref{isf2_flowchart}). Note that $f_2$ appears on both the right- and left-hand side of \autoref{eqn:marnewf2}, thus we no longer need to differentiate between odd and even iterations. Instead, the individual equation iteratively finds the full wavefield Green's and focusing functions. However, these functions are contaminated by the imperfect sampling, which needs to be deblurred using a PSF with each iteration. The first step in finding this PSF is estimating quantity $Y_{2,k}$ as follows (\autoref{eqnap:id2}):
\[
\delta(\xs{H,A}'-\xs{H,A})\delta(t)= 
\]
\begin{equation}
\hspace*{1cm} \int_{\mathbb{S}_0} Y_{2,k}(\xs{A}',\xs{S},t) *f_{2,k-1}(\xs{A},\xs{S},t)d\xs{S}.
\end{equation}
In this equation, $f_{2,k-1}$ is the deblurred version of the focusing function from the previous iteration. The focusing functions and its inverse are then used to approximate the PSF for the current iteration (\autoref{eqnap:gamma2plus}):
\[
\Gamma_{2,k}(\xs{A}',\xs{A},t)= 
\]
\begin{equation}
\hspace*{1cm} \sum_{i} Y_{2,k}(\xs{A}',\xs{S}^{(i)},t)* f_{2,k-1}(\xs{A},\xs{S}^{(i)},t).
\end{equation}
Subsequently, this PSF is used to deblur both the Green's and focusing functions from the step 2 of \autoref{isf2_flowchart} (Equations \ref{eqnap:marnewf2_G} and \ref{eqnap:marnewf2_f}). Finally, $-f_{2,k}^{\star}$ (the superscript $\star$ indicates time-reversal) is separated from the Green's function using a time-gate, reversed in time, and multiplied with $-1$ (step 6 in \autoref{isf2_flowchart}). This updated focusing function is then used as input for the next iteration. This process can then be repeated until the results have sufficiently converged, meaning the updated focusing function does not change significantly compared to the preceding iteration. \\
This new scheme successfully avoids the inversion of $f^-_1$, and it is therefore more stable. However, the drawback of this scheme is that it does not return the Green's function decomposed into an up- and down-going part. These decomposed functions are required for redatuming the reflection response from the surface to the focal depth. This issue can be circumvented, if either the downgoing or upgoing focusing functions is available. Previously, we found that the odd iterations in the flowchart of \autoref{is_flowchart} are relatively stable, thus this can be used to estimate $f^-_{1,k}$ and $G^-$ for odd iterations. The final step is now to find an approximation for $f^+_{1,k}$ in the even iterations, which can no longer use the even iterations of \autoref{is_flowchart}, since these iterations introduce the unstable inverse $Y_1$. Instead the relation in \autoref{eqn:deff2} is used to find the update for the downgoing focusing function, as follows:
\begin{equation}
\label{eqn:fpdecomp}
f^+_{1,k}(\xs{R},\xs{A},t) =  f_{2,k}(\xs{A},\xs{R},t) + f^-_{1,k-1}(\xs{R},\xs{A},-t) 
\end{equation}
Note that both $f^-_{1,k-1}$ as well as $f_{2,k}$ have already been deblurred, thus there are no sampling artifacts in $f^+_{1,k}$ (e.g. no PSF correction is required). Furthermore, \autoref{eqn:inif1p} is still valid to calculate $f^+_{1,0}$ for the initial iteration. Finally, the downgoing Green's function can be calculated after the last iteration using \autoref{eqn:greensdecomp}:
\begin{equation}
\label{eqn:greensdecomp2}
G^+(\xs{A},\xs{R},t)= G(\xs{R},\xs{A},t) - G^-(\xs{A},\xs{R},t).
\end{equation}

\begin{figure}
\centering
\includegraphics[trim={0.4cm 0cm 0cm 0.cm},clip,width=0.48\textwidth]{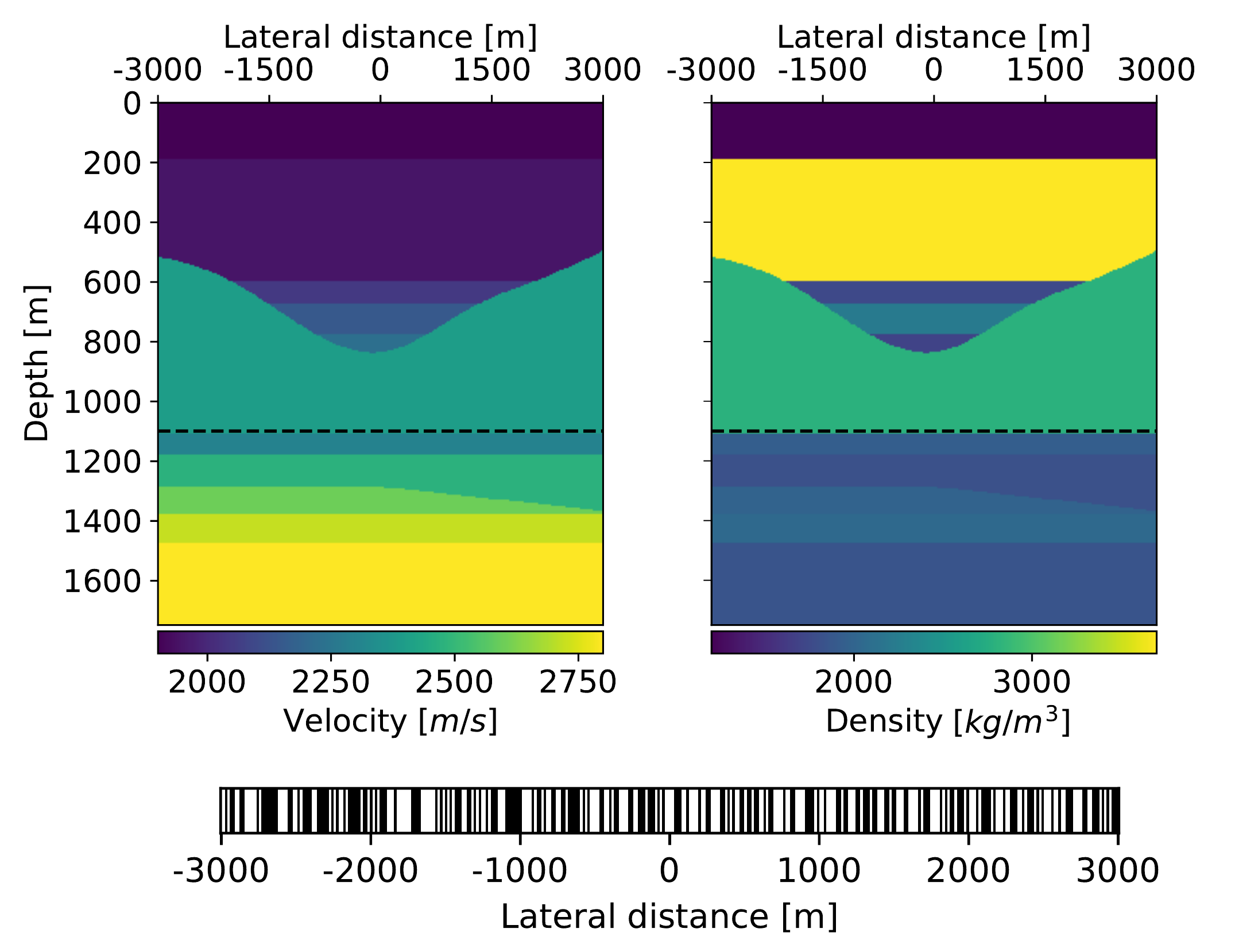} 
\caption{Model used in the numerical irregular sampling experiment, with the velocities on the left and densities on the right. The dashed line shows the focal level. The barcode shows the irregular sampling, with the white spaces denoting the excluded sources (50 \%).}
\label{isf2_model}
\end{figure}

\begin{figure*}
\centering
\includegraphics[width=\textwidth]{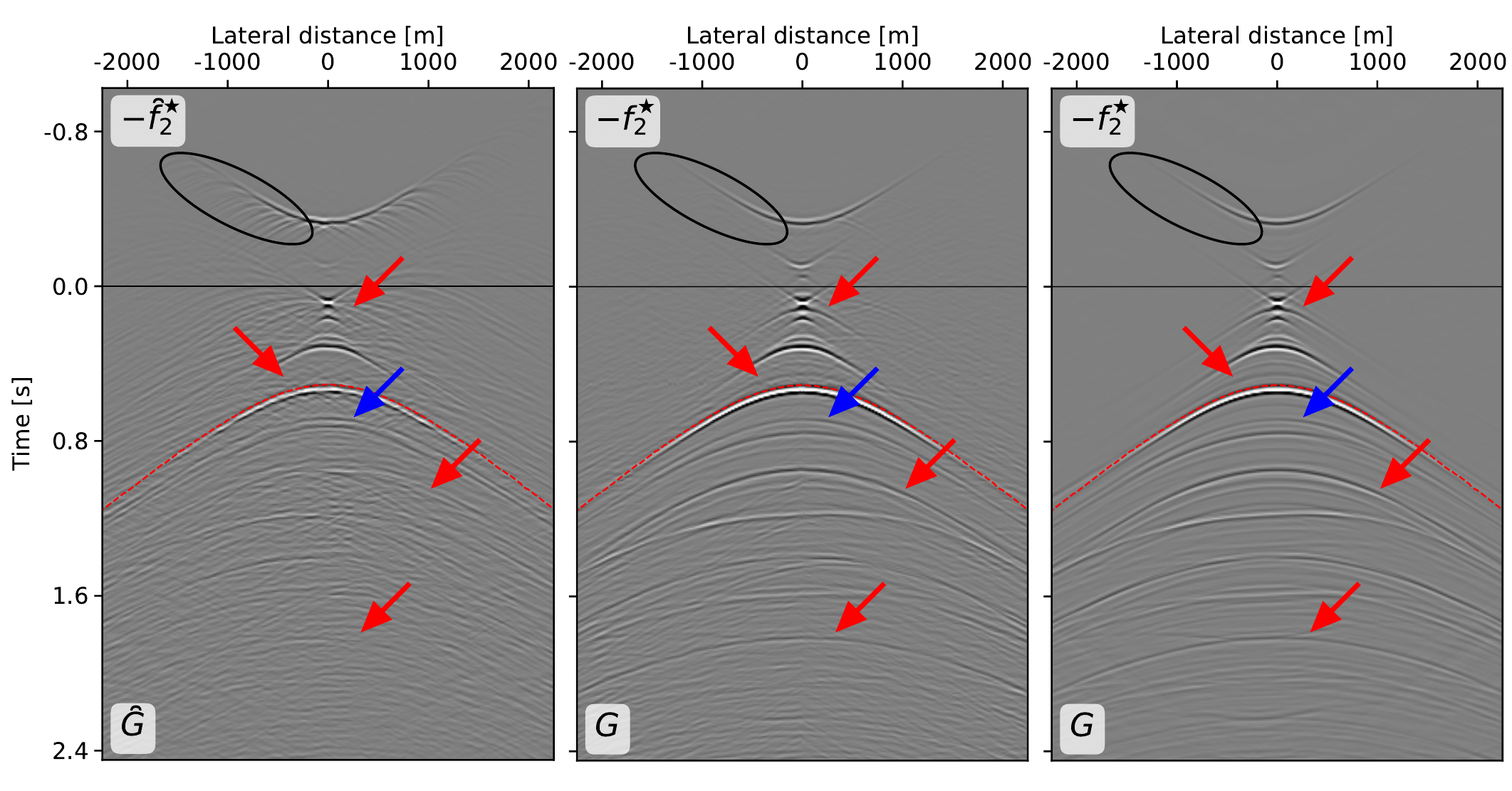} \\
\caption{The left panel shows the result of irregularly sampled data after 10 iterations of the standard Marchenko scheme. The middle panel shows the results when using our scheme on the same data, again 10 iterations are used. Finally, the 3rd panel shows the reference result, obtained after 10 iterations of the standard Marchenko scheme with well-sampled data. Each panel is scaled with its maximum value. The arrows and ellipses show artifacts arising from the irregular sampling. Distortions caused by the irregular sampling are indicated with the ellipses. The red arrows show events that deviate in amplitude or are missing altogether. Finally, the blue arrow marks an erroneous reflector. }
\label{isp_results}
\end{figure*}

\begin{figure*}
\centering
\includegraphics[width=\textwidth]{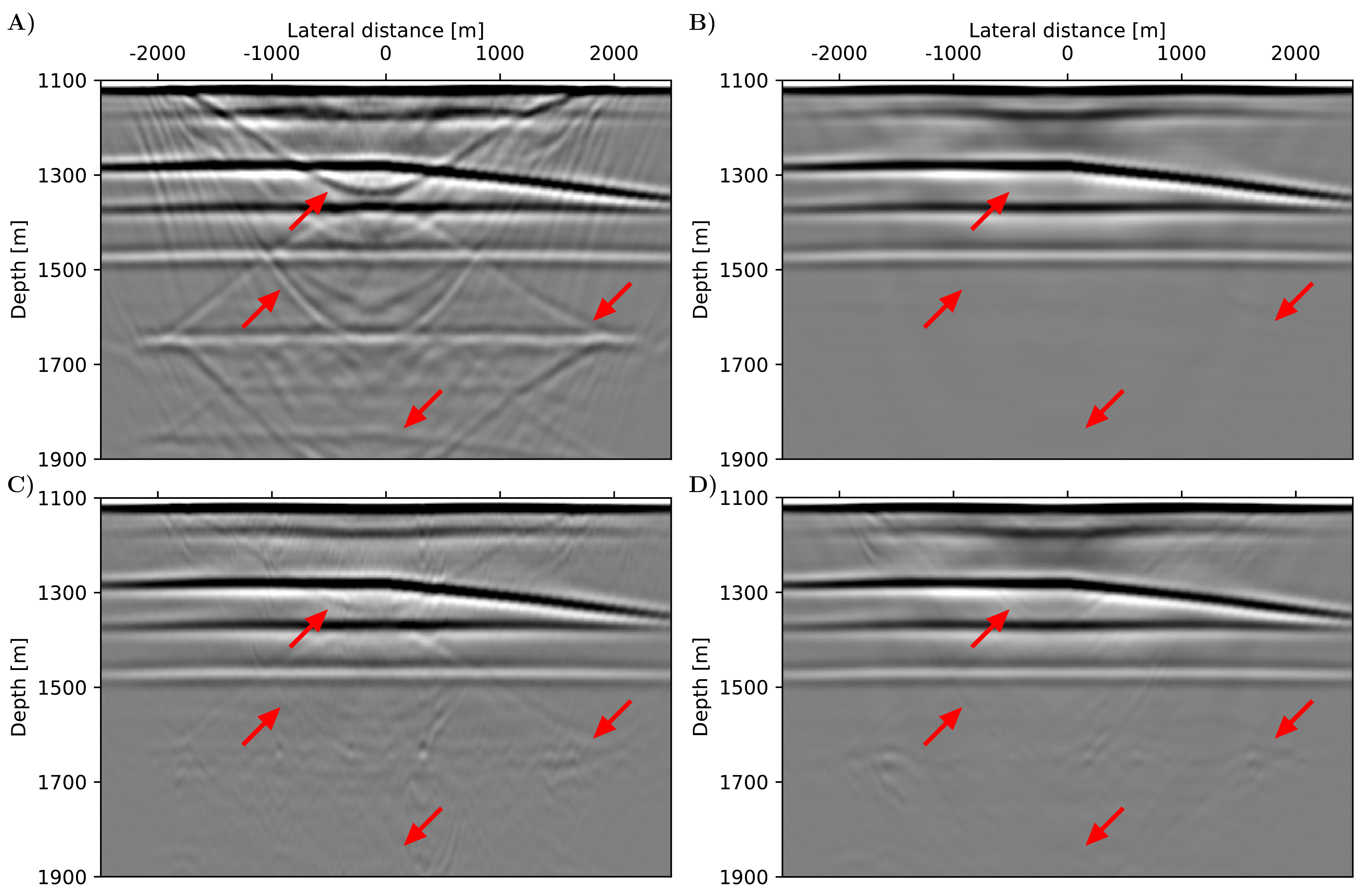} \\
\caption{Images of the target zone (i.e. at depths below 1100 meters). A) shows the migration of the redatumed reflection response, retrieved from the irregularly sampled data after 10 iterations of the standard Marchenko scheme. B) displays the reference migration, obtained using the results after 10 iterations of the standard Marchenko scheme with well-sampled data.. C) is the migration after 10 iterations of the newly proposed scheme on the same data. Finally, D) shows the results of reconstructing the reflection data first, and then applying 10 iterations of the standard Marchenko scheme.  Each panel is scaled with it's maximum value. The arrows show  overburden effects that are not completely eliminated due to the use of irregularly sampled data.}
\label{mig_results}
\end{figure*}
\noindent
From now on, the scheme introduced in this section and the scheme introduced before (summarized in \autoref{is_flowchart}) will be referred to as the full wavefield scheme and the decomposed scheme, respectively.



\section{full wavefield scheme numerical example}

Now, the full wavefield scheme will be tested with a numerical example. \autoref{isf2_model} shows the new velocity and density models that are used for this example. Contrary to the previous numerical example, there is no requirement for a strong contrast in acoustic impedance between the top two layers. The direct arrival of the Green's function is calculated in a smooth version of this model. The other parameters for modeling the reflection response remain the same, meaning that the source wavelet has a flat spectrum, and that 601 collocated sources and receivers are placed with a 10 meters separation. For the imperfect sampling, again 50\% of sources are removed, as depicted by the barcode plot in \autoref{isf2_model}. \\
\autoref{isp_results} presents the resulting Green's and focusing functions after 10 iterations of the full wavefield scheme. The first panel shows the results when using imperfectly sampled data with the standard full wavefield scheme (i.e. using only steps 1,2 and 6 in \autoref{isf2_flowchart}). Next, the middle panel shows the corrected Green's and focusing function obtained with the proposed full wavefield scheme (using all steps in \autoref{isf2_flowchart}). Finally, the third panel contains the reference result that is acquired with regularly sampled data. There are a number of interesting artifacts visible in the figure. First, sampling artifacts are highlighted by the black ellipse. Furthermore, the red arrows denote events that are not retrieved when using imperfectly sampled data. Lastly, the blue arrow marks a reflector recovered when using the imperfectly sampled data, that differs from the reflectors in the corrected and reference result. While a clear improvement can be observed when using the new full wavefield scheme, the PSF-corrected result still deviates considerably from the reference result. The match between the results especially deteriorates at larger times (i.e. at $t > 2s$). To further assess the performance of this method, the decomposed up- and down-going wavefields will now be considered, thus the full wavefield Green's function needs to be decomposed into the up- and down-going versions. As previously stated $G^-$ can be iteratively acquired by using the odd iterations in \autoref{is_flowchart} with \autoref{eqn:fpdecomp} to update $f^+$ for the even iterations. Subsequently, $G^+$ can be calculated from \autoref{eqn:greensdecomp2}. Using the decomposed Green's functions, the redatumed reflection response at the focal level can now be found, by means of the following relation \citep{wapenaar2014marchenko}:
\[
G^-(\xs{A},\xs{R},t) = 
\]
\begin{equation}
\hspace*{1cm}\int_{\mathbb{S}_0} R(\xs{A},\xs{A}',t) * G^+(\xs{A}',\xs{R},t) d\xs{A}'
\end{equation}
The redatumed reflection response is acquired from this equation with a MDD. Next, this reflection response is migrated, to get an image of the target, free from multiples related to the overburden. Note that this requires a smooth version of the velocity model below the focal depth. The results of the migration are displayed in \autoref{mig_results}. Panel A) and C) show the results using irregularly sampled data of the full wavefield scheme with and without PSF-correction, respectively. The panel in B) holds the reference Marchenko result obtained with regularly sampled data. Finally, D) shows a migration of results by the standard Marchenko scheme, where the irregularly sampled reflection data is reconstructed before applying the scheme. In order to achieve this reconstruction, a slight NMO correction is first applied to compress the range of ray parameters. Next, a sparse inversion using the Radon transform is used to restore the missing data. Lastly, the NMO correction is undone, and the originally available sources are combined with the reconstructed result to acquire the reconstructed reflection response. Even though the internal multiples are not perfectly suppressed using the PSF-corrections (\ref{mig_results}C), the result matches the reference image (\ref{mig_results}B) significantly better than the image without any corrections. The results with reconstructed reflection data (\ref{mig_results}D) realize an even better match with the reference.


\section{Discussion}

The results show that the proposed schemes can successfully be used on irregularly sampled reflection data. However, there are some limitations and possible improvements that will now discussed. \\ 
First, we note that the discretizations in \ref{eqn:dis1} and \ref{eqn:dis2} should be multiplied with the irregular integration step $\Delta \xs{S}^{(i)}$. However, the current implementation with PSFs uses a regular integration step ($\Delta \xs{S}=10m$) based on the regular grid of sources and receivers. This poses no issues for the schemes that apply the PSF-correction, as they implicitly correct for the irregular source distances.  
Nevertheless, one could argue that the irregular scheme without PSFs should include the irregular source distances instead of the regular distances. This approach was also tested, but did not significantly alter the results of the blurred images. \\
Second, the largest limitation when using the decomposed equations is the instability of quantity $Y_1$, which was introduced as the inverse of the upgoing focusing function. This was circumvented with the introduction of a full wavefield scheme that avoids this inverse. However, the full wavefield scheme appears to have decreased accuracy at later times, as observed when comparing the second and third columns of \autoref{is_results} and \autoref{isp_results}. \\
Another important factor is the computational cost of the method. For every iteration, the decomposed scheme adds one convolution and two MDD steps to the standard Marchenko scheme, which only consists of a single convolution per iteration. Furthermore, additional operations are required to decompose the results of the full wavefield scheme into up- and down-going responses. Specifically, the upgoing Green's and focusing functions have to be computed, according to the odd iterations of the decomposed scheme. Therefore, the full wavefield scheme adds a convolution and two inversions to the computational load for each iteration of the method, thus increasing the computational costs and time of the method. \\
Alternatively to the full wavefield scheme, the inversion of $f^-_1$ can also be avoided by utilizing a Marchenko scheme for data that include free-surface multiples \citep{singh2015}. This scheme would have the same number of operations as the decomposed scheme, and thus would come at a lower computational cost than the full wavefield scheme. However, including free-surface multiples can lead to instabilities in the Marchenko series \citep[e.g.][]{staring2017,dukalski2017}, but these instabilities are expected to be less troublesome than the instability of quantity $Y_1$. Nevertheless, further research is required to assess the viability of such a scheme. \\
Although, the new formulation no longer requires collocation of the sources and receivers in the Marchenko scheme, it is important to note that co-depthening is still required. Traditionally, this is achieved by redatuming the sources down to the receiver level after applying surface related multiple elimination \citep[SMRE,][]{verschuur1992SMRE}. However, SRME will also suffer from irregular acquisition effects, so a different scheme for removing free-surface multiples is desirable, such as estimating primaries by sparse inversion \citep[EPSI,][]{groenestijn2009EPSI}, which is less sensitive to the acquisition geometry. \\
While the inverse of the downgoing focusing function always exists, there is a different way to estimate the transmission response, which does not require any explicit inversions \citep{vasconcelos2018}. This methodology was also tested to calculate the transmission response in step 3 of the proposed decomposed scheme. While this method achieved promising results in 1.5D media, we found that the results were unsatisfactory in the 2D model. Therefore, the transmission response was estimated by inversion instead. \\
The new methodology is unable to account for irregular sampling of both sources and receivers; the sampling can only be irregular in the same dimension as the integration in equations \ref{eqn:mar1} and \ref{eqn:mar2}. On the contrary, the method introduced by \citet{haindl2018sparsity} assumes irregular sampling in the opposite dimension. A combination of these complementary methods is, therefore, envisioned to deal with irregular sampling in both the source and receiver dimensions simultaneously. However, further research into this topic is required. \\
Finally, we note that the reflection data can also be reconstructed before applying the Marchenko method. Subsequently, this interpolated reflection response can be used in the standard iterative scheme, as shown in \autoref{mig_results}. Although previous studies found that the resulting Green's and focusing functions contained a relatively high level of noise \citet{haindlmsc}, we demonstrate that careful reconstruction of the data can allow for accurate images of the target area. Moreover, these results show less artifacts than the PSF-driven full wavefield Marchenko scheme. The additional pre-processing, however, had a larger computational costs than the proposed full wavefield scheme (e.g. the method with reconstruction beforehand took approximately 24 hours on a single CPU, compared to 3 hours for the full wavefield Marchenko scheme).

\section{Conclusion}

One of the restrictions of the Marchenko method is the need for well-sampled and collocated sources and receivers. Recent work introduced new representations for irregularly sampled data. These representations include point-spread functions (PSFs) that deblur distorted focusing- and Green's functions. Based on these representations, this paper shows that the iterative Marchenko scheme can be adapted to handle irregularly sampled data. For this adaptation the location of the missing sources needs to be known, and an inverse version as opposed to the time-reversed version of the direct arrival of the Green's function is required as initial estimate of both new schemes. In addition, each iteration of the standard Marchenko scheme is extended by three steps. First, an approximation of the transmission response or quantity $Y_1$ needs to be computed for the odd and even iterations, respectively. Quantity $Y_1$ is the inverse of the upgoing focusing function, similar as the transmission response is the inverse of the downgoing focusing function. Second, these approximations are irregularized in accordance with the missing sources. Subsequently, these irregular versions are used to calculate a PSF. Third, the well-sampled focusing- and Green's functions are reconstructed by a multidimensional deconvolution of the blurred original functions with these PSFs. \\
While the decomposed scheme shows promising initial results, it is established that quantity $Y_1$ is not necessarily stable. Therefore, a second full wavefield scheme is proposed, which does not rely on the unstable $Y_1$. This is achieved by combining the two decomposed equations into a single full wavefield equation. This also yields a new iterative full wavefield scheme, which analogous to the first decomposed scheme contains three additional steps compared to the classical Marchenko scheme. Again, these steps resolve and apply a PSF to correct for imperfect sampling in the retrieved responses. A numerical example shows that the full wavefield scheme succeeds in suppressing internal multiples in the final Marchenko image, whereas the classical approach fails to eliminate the internal multiples when imperfectly sampled data is used. \\
The newly proposed schemes alleviate the need for well-sampled sources when using the Marchenko method. Ideally, the need for well-sampled receivers should be removed as well. While this is subject to ongoing research, a new scheme involving a sparse inversion is envisioned. By relaxing the need for perfectly sampled data, the Marchenko method can be more widely applied to field data.

\begin{acknowledgments}
The authors thank Matteo Ravasi and Dominic Cummings for their insightful reviews of the manuscript. We are also grateful to Jan Thorbecke, Christian Reinicke, Eric Verschuur, Joeri Brackenhoff and Max Holicki for help with the numerical examples and insightful discussions. This research was funded by the European Research Council (ERC) under the European Union's Horizon 2020 research and innovation programme (grant agreement No: 742703).
\end{acknowledgments}

\bibliographystyle{gji}
\bibliography{Irr_bib}

\appendix
\section{Derivation of irregular full wavefield scheme} 
\setcounter{equation}{0}
\renewcommand{\theequation}{A\arabic{equation}}

This appendix proposes new representations for irregular sampling in the full wavefield Marchenko scheme. This full wavefield scheme is used instead of the decomposed Marchenko equations, to avoid the use of the unstable inverse of $f_1^-$. First, \autoref{eqn:mar1} and \autoref{eqn:mar2} are combined to get a single Marchenko representation for the full wavefield Green's function, giving \citep{wapenaar2014marchenko}:
\[
G(\xs{R},\xs{A},t) - f_2(\xs{A},\xs{R},-t) = 
\]
\begin{equation}
\label{eqnap:marf2}
\hspace*{1cm} \int_{\mathbb{S}_0} R(\xs{R},\xs{S},t) * f_2(\xs{A},\xs{S},t)d\xs{S},
\end{equation}
with:
\begin{equation}
\label{eqnap:deff2}
f_2(\xs{A},\xs{R},t) =
 f^+_1(\xs{R},\xs{A},t) -  f^-_1(\xs{R},\xs{A},-t).
\end{equation}
Similarly as with the decomposed schemes, the right-hand side integral in \autoref{eqnap:marf2} is approximated by finite summations over the available sources:
\begin{equation}
\label{eqnap:disf2}
\sum_{i} R(\xs{R},\xs{S}^{(i)},t) * f_2(\xs{A},\xs{S}^{(i)},t)*S(t).
\end{equation}
The discretization in \autoref{eqnap:disf2} is the source of the distortions in the case of an imperfectly sampled reflection response. \\
The next objective is to find a new PSF, that will correct for these distortions. Again, we utilize the fact that a convolution of the focusing focusing with it's reverse produces a band-limited delta pulse. We define response $Y_2$ as the inverse of $f_2$ in \autoref{eqnap:deff2}, as follows: 
\[
\delta(\xs{H,A}'-\xs{H,A})\delta(t)= 
\]
\begin{equation}
\label{eqnap:id2}
\hspace*{1cm} \int_{\mathbb{S}_0} Y_2(\xs{A}',\xs{S},t)* f_2(\xs{A},\xs{S},t)d\xs{S}.
\end{equation}
Alternatively we find:
\[
\delta(\xs{H,S}-\xs{H,S}')\delta(t)= 
\]
\begin{equation}
\label{eqnap:idalt2}
\hspace*{1cm} \int_{\mathbb{S}_A} f_2(\xs{A},\xs{S},t) * Y_2(\xs{A},\xs{S}',t)d\xs{A}.
\end{equation}
We note that this inverse more stable than inverse $Y_1$ of $f_1^-$, because of the presence of $f_1^+$ in the definition of $f_2$. Again the irregular sampling is applied to the integral, resulting in a summation over the irregular sources:
\[
\Gamma_2(\xs{A}',\xs{A},t)= 
\]
\begin{equation}
\label{eqnap:gamma2plus}
\hspace*{1cm} \sum_{i} Y_2(\xs{A}',\xs{S}^{(i)},t)* f_2(\xs{A},\xs{S}^{(i)},t)*S(t).
\end{equation}
\autoref{eqnap:gamma2plus} is the new PSF for the full wavefield Marchenko representations. This PSF is convolved with the right-hand side of \autoref{eqnap:marf2}:
\[
\int_{\mathbb{S}_A} \int_{\mathbb{S}_0} R(\xs{R},\xs{S},t) 
\]
\begin{equation}
\hspace*{1cm} * f_2(\xs{A}',\xs{S},t)*\Gamma_2(\xs{A}',\xs{A},t)d\xs{S}d\xs{A}'.
\end{equation}
Next, the order of integration and summation is reversed, and  we find, using \autoref{eqnap:gamma2plus}, as well as \autoref{eqnap:idalt2}, and the sifting property of the delta function:
\[
\sum_{i}  \int_{\mathbb{S}_0} R(\xs{R},\xs{S},t) * 
\]
\[
\hspace*{1cm} \int_{\mathbb{S}_A} f_2(\xs{A}',\xs{S},t)  * Y_2(\xs{A}',\xs{S}^{(i)},t) d \xs{A}'  
\]
\[
\hspace*{1cm} * f_2(\xs{A},\xs{S}^{(i)},t)*S(t) d\xs{S} = 
\]
\[
\sum_{i} \int_{\mathbb{S}_0} R(\xs{R},\xs{S},t) * 
\delta(\xs{H,S}-\xs{H,S}^{(i)}) \delta(t)d\xs{S}  
\]
\[
\hspace*{1cm}  * f_2( \xs{A},\xs{S}^{(i)},t)*S(t) = 
\]
\begin{equation}
\sum_{i} R(\xs{R},\xs{S}^{(i)},t) * f_2(\xs{A},\xs{S}^{(i)},t)*S(t).
\end{equation}
Note that this is identical to \autoref{eqnap:disf2}. Finally, $\int_{\mathbb{S}_A} \{\cdot\} * \Gamma_2 d\xs{A}'$ is applied to both sides of \autoref{eqnap:marf2}, giving:
\[
\invbreve{G}(\xs{R},\xs{A},t) - \invbreve{f}_2(\xs{A},\xs{R},-t) =
\]
\begin{equation}
\label{eqnap:marnewf2}
\hspace*{1cm} \sum_{i} R(\xs{R},\xs{S}^{(i)},t) * f_2(\xs{A},\xs{S}^{(i)},t)*S(t),
\end{equation}
with:
\[
\invbreve{G}(\xs{R},\xs{A},t) = 
\]
\begin{equation}
\label{eqnap:marnewf2_G}
\hspace*{1cm} \int_{\mathbb{S}_A} G(\xs{R},\xs{A}',t) * \Gamma_2(\xs{A}',\xs{A},t) d\xs{A}',
\end{equation}
and
\[
\invbreve{f}_2(\xs{A},\xs{R},-t) = 
\]
\begin{equation}
\label{eqnap:marnewf2_f}
\hspace*{1cm} \int_{\mathbb{S}_A} f_2(\xs{A}',\xs{R},-t) * \Gamma_2(\xs{A}',\xs{A},t) d\xs{A}'.
\end{equation}

\bsp

\balance

\end{document}